\documentclass[namedreferences]{SolarPhysics}
\usepackage[optionalrh,solaenum]{spr-sola-addons} 
\usepackage{graphicx}                    
\usepackage{color}                       
\usepackage{url}                         

\usepackage{graphics}
\usepackage{epsfig} 
\usepackage{subfig}
\usepackage{caption}
\usepackage{lscape}
\usepackage{multirow}
\usepackage{multicol}
\usepackage{ctable}
\usepackage{rotating}
\usepackage{amsfonts}
\usepackage{fixltx2e}
\usepackage[flushleft]{threeparttable}

\newcommand{\aap}{Astron. Astrophys.}
\newcommand{\apj}{Astrophys. J.}
\newcommand{\solphys}{Solar Phys.}

\newcommand{\ssr}{Space Sci. Rev.}

\newcommand{\apjl}{Astrophys. J. Lett.}

\newcommand{\sw}{Space Weather}

\begin{document}

\begin{opening}

\title{Exploring the capabilities of the Anti-Coincidence Shield of the INTEGRAL spectrometer to study solar flares
\\ {\it Solar Physics}}

%
\author{R.~\surname{Rodr\'{\i}guez-Gas\'{e}n}$^{1,2}$\sep
              J.~\surname{Kiener}$^{1}$\sep              
              V.~\surname{Tatischeff}$^{1}$\sep
              N.~\surname{Vilmer}$^{2}$ \sep
              C.~\surname{Hamadache}$^{1}$\sep
              K.-L.~\surname{Klein}$^{2}$}	     


\runningauthor{Rodr\'{\i}guez-Gas\'{e}n et al.}
\runningtitle{The Anti-Coincidence Shield of the INTEGRAL spectrometer to study solar flares}


    \institute{$^{1}$ CSNSM, IN2P3-CNRS, Univ. Paris-Sud, F-91405 Orsay Cedex, France\\ 
                    $^{2}$ LESIA, Observatoire de Paris, CNRS, UPMC and Univ. Paris-Diderot, F-92195 Meudon Cedex, France. Email: \url{rosa.rodriguez@obspm.fr}}

\begin{abstract}
INTEGRAL is a hard X-ray/$\gamma$-ray observatory for astrophysics (ESA) covering photon energies from 15~keV to 10~MeV. It was launched in 2002 and since then the BGO detectors of the Anti-Coincidence shield (ACS) of the SPI spectrometer have detected many hard X-ray (HXR) bursts from the Sun, producing lightcurves at photon energies above $\sim$100~keV. The spacecraft has a highly elliptical orbit, providing a long uninterrupted observing time (about 90\% of the orbital period) with nearly constant background due to the reduction of the crossing time of the Earth's radiation belts. However, due to technical constraints, INTEGRAL cannot point to the Sun and high-energy solar photons are always detected in non-standard observation conditions. To make the data useful for solar studies, we have undertaken a major effort to specify the observing conditions through Monte-Carlo simulations of the response of ACS for several selected flares. We check the performance of the model employed for the Monte-Carlo simulations using RHESSI observations for the same sample of solar flares. We conclude that, despite the fact that INTEGRAL was not designed to perform solar observations, ACS is a useful instrument in solar flare research. In particular, its relatively large effective area allows the determination of good-quality HXR/$\gamma$-ray lightcurves for X- and M-class solar flares and, in some cases, probably also for C-class flares.
\end{abstract}

\keywords{X-Ray Bursts, Association with Flares; Flares, Energetic Particles; Instrumental Effects}
\end{opening}
\section{Introduction}\label{s1}

Solar flares are the most powerful explosion events in the solar system, able to release, in tens of minutes, up to 10$^{32}$-10$^{33}$~ergs of energy. They and their (sometimes) associated fast coronal mass ejections (CMEs) are high energy particle accelerators, generating ions up to tens of GeV and electrons to hundreds of MeV (\opencite{Ramaty95}; \opencite{Ramaty00}; \opencite{Hudson04}). Solar flares emit radiation across the entire EM spectrum, from radio to $\gamma$-ray. Flare-accelerated electrons and ions interacting with the ambient solar atmosphere produce bremsstrahlung HXR/$\gamma$-ray continuum \cite{Lin06,Holman11} and $\gamma$-ray line emission \cite{Ramaty86,Vilmer03,Share06}, respectively. Quantitative information about the parent distributions of electrons and ions can be inferred from HXR \cite{Kontar11} and $\gamma$-ray line measurements \cite{Vilmer11}, which introduce strong constraints on the acceleration mechanisms and interaction regions.

In particular, knowledge of the HXR/$\gamma$-ray photon spectrum provides fundamental information about the electron/ion beam spectra, distribution and energy contents \cite{Vilmer12}. HXR/$\gamma$-ray observations obtained with high spectral resolution are now achievable with RHESSI \cite{Lin02} and INTEGRAL \cite{Winkler03} spacecraft (see for example, \opencite{Lin03}; \opencite{Hurford03}; \opencite{Kiener06}). Exploiting a general purpose satellite such as INTEGRAL for solar observations is useful, because of the necessity of more observations due to the incomplete coverage of the spacecraft aimed at it. Such exploitation, however, needs detector simulations in order to identify the response of the instrument in the context of solar observations, and to correct the count rates for the detector geometry relative to the Sun, which varies from event to event. This is the scope of the present paper.

INTEGRAL spacecraft was designed to get fine spectroscopy and fine imaging of celestial (galactic and extragalactic) $\gamma$-ray sources in the energy range of 15~keV to 10~MeV, with concurrent source monitoring in the X-ray and optical energy ranges. The SPI spectrometer \cite{Vedrenne03}, with its main component being the germanium (Ge) detector matrix,  observes $\gamma$-rays between 20~keV and 8~MeV with an energy resolution of 2.5~keV at 1.33~MeV. Its Anti-Coincidence Shield (ACS) is one of the largest detectors of astrophysical HXR and $\gamma$-ray photons presently in orbit, and can be used to get lightcurves without energy information. ACS consists in 91 different BGO\footnote{~Bismuth Germanate.} scintillator crystals of various shapes surrounding the Ge-matrix of SPI. They are configured into a cylinder around SPI (an upper and lower collimator ring), and a side and rear shield. The collimator rings, together with the tungsten coded mask, define the field-of-view of SPI. 

It is important to point out the fact that ACS does not record energy information. It provides count rates with 50 ms time resolution above some photon energy threshold in the range of some tens of keV. This limitation may be overcome from modelling of the instrument. The effective energy value, for example, depends on the viewing angle between the spacecraft axis, the photon source (here, the Sun), as well as on the electronic threshold of the BGO-detectors. This information may be derived by means of Monte-Carlo simulations. Another example is the response of the instrument to incoming photon beams with different spectral indexes, that may also be simulated.

INTEGRAL was designed so that IBIS (the imager instrument aboard INTEGRAL) must be always between SPI and the Sun, to provide shadowing to the SPI cooling system. SPI can never observe the Sun directly (it is typically to the rear of the instrument), and solar photons enter the instrument through the side or backside. Despite this fact, the Ge matrix of SPI located inside the ACS can also be reached, and $\gamma$-ray spectra in the 1-10~MeV range may be available. The lightcurves and spectra of three X-class solar flares in the late phase of solar cycle 23 have been extracted and analyzed \cite{Gros04,Kiener06,Harris07}, so far.

The work presented here is carried out within the European FP7 project SEPServer (\opencite{Vainio13}; \opencite{Malandraki12}). The main aim of the SEPServer project is to build an on-line server that will provide the space research community with solar energetic particle (SEP) data and related observations of solar EM emission, with access to scattered data from a number of international observatories. Under this project, we are recollecting HXR/$\gamma$-ray data recorded by INTEGRAL for a set of solar flares (related to SEP events) that occurred during the 23rd solar cycle (1996-2006). By means of Monte-Carlo simulations, we are providing information about the ACS effective area for different incoming photon energies, different spacecraft pointing configurations (orientations of the spacecraft with respect to the Sun) and varying the electronic threshold of the BGO detectors, which is not completely known. We are also giving information about the solar observing conditions of the data under study, as well as the effective area that would characterize each solar flare. Furthermore, we are comparing observational ACS data with the expected number of counts obtained from the Monte-Carlo simulations, when using RHESSI data and the simulated response of the BGO detectors. 

The organization of the article is as follows: in Section~\ref{s2} we present the data selection considered for this study together with the observational data for the case of the June 17, 2003 solar flare (as a matter of example). Section~\ref{s3} explains the Monte-Carlo code used for the simulations, a brief study about its performance, the initial parameters used for the simulations and the observational conditions for each solar flare. Section~\ref{s4} contains the results obtained with the simulations, together with the comparison of observational ACS data with those expected. Section~\ref{s5} summarizes and concludes the work.

\begin{landscape}
\begin{table}
\caption{List of solar flares. [1] date; [2] SXR peak time; [3] SXR class; [4] H$_{\alpha}$ location; [5] spacecraft azimuthal angle; [6] spacecraft meridional angle; [7] time interval of ACS detection; [8] number of ACS counts; [9] effective area for photons with $E$ $>$ 50~keV and a initial power law with spectral index $\alpha$ = 3 (see text). Notes : (a) noteworthy, the HXR/$\gamma$-ray detection of this flare by ACS started significantly after the recorded peak time of the SXR (GOES). The origin of such discrepancy is not known; (b) due to fluctuating background, we adopted the same time interval as for the RHESSI detection (A. Shih, private communication); (c) very high energy flux: ACS saturation at 07:00:25~UT; (d) the number of counts is given without ACS dead-time correction due to Ge-matrix annealing during this period (prepared scientific data is not available, only raw data from the BGO cristals).}
\label{table1}
\begin{tabular}{lcccccccc}
\hline
Date & Time & Class & H$_{\alpha}$& $\theta_{\rm obs}$ & $\varphi_{\rm obs}$ & $\Delta t_{\rm ACS}$ & $N_{\rm obs}({\rm ACS})$ & {A}$_{\rm eff}$\\
 & [UT] &  &  &  [$^{\circ}$] & [$^{\circ}$] & [hh:mm:ss] & & [cm$^{2}$] \\
\hline
\textbf{2002} & & & & & & \\
2002 Nov 09 & 13:23 & M4.9 & S12W29 & 75.4 & -0.5 & 13:12:40~-~13:28:00 & $(9.4 \pm 0.4)\times 10^5$ & 423$^{+80}_{-151}$\\
2002 Dec 19 & 21:53 & M2.7 & ? & 66.9 & 0.1 & 21:38:40~-~21:46:30 & $(1.2 \pm 0.1)\times 10^5$ & 528$^{+120}_{-206}$\\
\textbf{2003} & & & & & & \\
2003 Apr 23 & 01:06 & M5.1 & N22W25 & 117.0 & 0.3 & 01:01:20~-~01:03:00 & $(1.1 \pm 0.1)\times 10^5$ & 217$^{+71}_{-119}$\\
2003 May 28& 00:27 & X3.9 & S07W21 & 96.3 & 0.5 & 00:21:15~-~00:38:50 & $(6.3 \pm 0.1)\times 10^6$ & 237$^{+55}_{-95}$\\
2003 May 31 & 02:24 & X1.0 & S07W65 & 121.2 & -1.6 & 02:18:50~-~02:30:05 & $(2.1 \pm 0.1)\times 10^6$ & 226$^{+79}_{-129}$\\ 
2003 Jun 17 & 22:45 & M6.8 & S08E61 & 75.8 & -0.3 & 22:45:00~-~23:10:00 & $(7.3 \pm 0.3)\times 10^6$ & 419$^{+79}_{-149}$\\
2003 Oct 28 & 11:10 & X18.4S & S16E07 & 122.0 & 0.7 & 11:01:40~-~11:25:00 & $(1.1 \pm 0.1)\times 10^8$ & 224$^{+77}_{-129}$\\
2003 Nov 2 & 17:25 & X9.3 & S14W56 & 75.4 & 0.6 & 17:15:00~-~17:29:30 & $(5.6 \pm 0.2)\times 10^7$ & 420$^{+79}_{-151}$\\
2003 Nov 4 & 19:44 & X18.4S & S19W83 & 126.3 & 0.1 & 19:40:10~-~20:00:00 & $(9.9 \pm 0.2)\times 10^7$ & 250$^{+79}_{-141}$\\
\textbf{2004} & & & & & & \\
2004 Apr 11 & 04:19 & M1.0 & S14W47 & 72.8 & 0.7 &04:13:50~-~04:15:50 & $(2.6 \pm 0.1)\times 10^5$ & 462$^{+95}_{-172}$\\
2004 Jul 13 & 00:17 & M6.7 & N14W45 & 111.1 & 0.2 & 00:13:20~-~00:17:45 & $(4.6 \pm 0.1)\times 10^5$ & 183$^{+51}_{-93}$\\ 
2004 Nov 7 $^{(a)}$ & 16:06 & X2.2 & N09W17 & 100.1 & 3.1 & 16:24:45~-~17:02:45 & $(2.0 \pm 0.2)\times 10^6$ & 217$^{+59}_{-95}$\\
2004 Nov 9 & 17:19 & M8.9 & N08W51 & 94.2 & -2.9 & 17:07:00~-~17:18:10 & $(6.8 \pm 0.5)\times 10^5$ & 290$^{+79}_{-123}$\\
2004 Nov 10 & 02:13 & X2.8 & N09W49 & 94.7 & 3.1 & 02:06:40~-~02:15:00 & $(2.8 \pm 0.10)\times 10^6$ & 285$^{+86}_{-122}$\\
\textbf{2005} & & & & & & \\
2005 Jan 17 $^{(b)}$ & 09:52 & X4.2 & N14W24 & 111.9 & -2.8 & 09:42:20~-~10:08:10 & $(1.0 \pm 0.4)\times 10^7$ & 190$^{+58}_{-99}$\\
2005 Jan 20 $^{(c)}$ & 07:00 & X7.9 & N12W58 & 118.8 & -0.3 & 06:40:00~-~07:00:25 &  $>1.0\times 10^{10}$ & 229$^{+78}_{-129}$\\
2005 May 13 & 16:57 & M8.5 & N12E11 & 117.1 & -0.4 & 16:36:35~-~16:54:20 & $(5.5 \pm 0.1)\times 10^5$ & 214$^{+74}_{-119}$\\
2005 Jul 13 & 14:49 & M5.6 & N13W75 & 67.6 & 1.8 & 13:53:00~-~14:33:50 & $(3.7 \pm 0.2)\times 10^6$ & 525$^{+121}_{-206}$\\
2005 Jul 13 & 21:54 & M1.2 & N08W90 & 66.2 & 0.2 & 21:51:40~-~21:53:25 & $(4.4 \pm 0.1)\times 10^5$ & 539$^{+122}_{-212}$\\
2005 Sep 7 & 17:40 & X18.1 & S06E89 & 77.9 & 0.1 & 17:26:55~-~17:59:50 & $(2.0 \pm 0.1)\times 10^8$ & 406$^{+77}_{-146}$\\
2005 Sep 13 & 19:27 & X1.5 & S09E05 & 115.5 & 0.8 &  19:14:05~-~19:17:40 & $(1.7 \pm 0.1)\times 10^6$ & 215$^{+68}_{-116}$\\
\textbf{2006} & & & & & & \\
2006 Jul 6 & 08:36 & M2.5 & S09W34 & 65.1 & -0.5 & 08:22:20~-~08:24:40 & $(5.3 \pm 0.5)\times 10^4$ & 544$^{+127}_{-215}$\\
2006 Dec 13  $^{(d)}$ & 02:40 & X3.4 & S06W23 & 95.1 & -0.2 & 02:21:20~-~02:34:25 &  $(2.0 \pm 0.2)\times 10^6$ & 240$^{+56}_{-95}$\\
2006 Dec 14  $^{(d)}$ & 22:15 & X1.5 & S06W46 & 86.2 & -0.1 & 22:05:45~-~22:29:20 & $(1.2 \pm 0.3)\times 10^6$ & 333$^{+76}_{-125}$\\
\hline
\end{tabular}
\end{table} 
\end{landscape}
\section{Observing solar flares with INTEGRAL}\label{s2}

Based on the list published by \inlinecite{Laurenza09} and the catalogue generated for SEPServer \cite{Vainio13}, Table~\ref{table1} gives the characteristics and the results of the Monte-Carlo simulations of those solar flares considered in this study. For all these cases, INTEGRAL/ACS detected a solar burst. From left to right, columns display [1] the date of the flare occurrence, [2] the peak time of the Soft-X rays (SXR), [3] the flare class (obtained from GOES measurements) and [4] its location (from H$_{\alpha}$ observations). Columns [5]-[6] list the coordinates of INTEGRAL used for the simulations, [7] the time interval of solar photon detection with ACS, [8] the observed number of ACS counts, and [9] the ACS effective area for each event, as deduced from the Monte-Carlo simulations (see below). 

ACS data were analyzed for 45 solar flares of the SEPServer catalog that occurred between October 2002 (INTEGRAL launch date) and December 2006. In each lightcurve, we looked for a statistically significant count excess ($> 5\sigma$ above background expectation) during more than one minute in a time interval of $\pm 20$~min around the SXR peak time (Table~1, column [2]). These criteria led to the detection of 24~solar events. The number of ACS counts for each event was obtained after subtraction of a background count rate estimated from a time interval of 10~min just before the onset of the flare. For some long-duration events, e.g. for the solar flare of October 28, 2003, we have considered that the background radiation increased before the end of the hard X-ray flare due to the arrival of SEPs. The uncertainty in background subtraction was added quadratically to the (small) statistical errors in the number of ACS counts (see Table~1).

\begin{figure}[!h]
\centering
\includegraphics[width=1.0\textwidth]{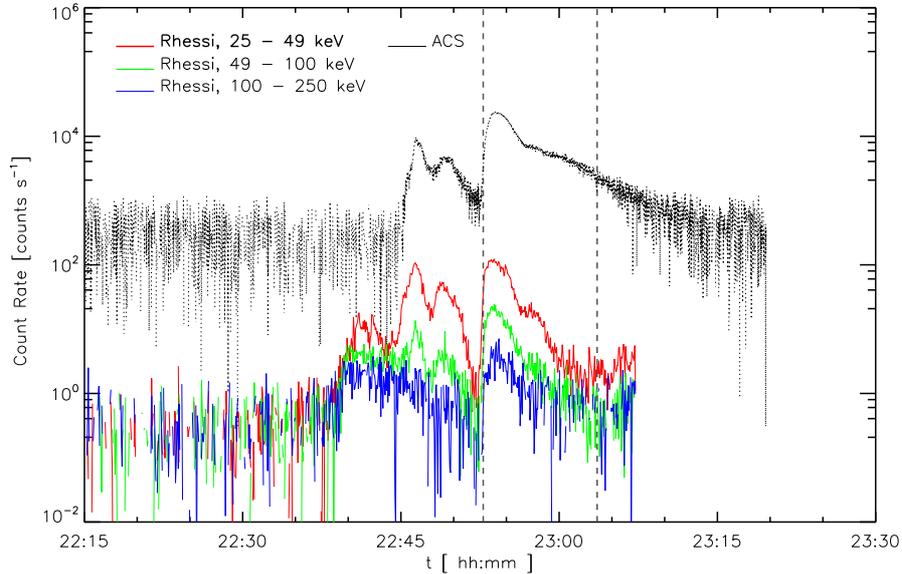}
\caption{Background subtracted ACS (in black) and RHESSI count rate data for three energy channels (25\,-\,49\,keV in red, 49\,-\,100\,keV in green and 100\,-\,250\,keV in blue) of the June 17, 2003 solar flare. The two dashed vertical lines indicate the time interval considered to compute the counts recorded by ACS for this flare to be compared with RHESSI data (see text and section~\ref{s4.2}).}
\label{ACS}
\end{figure}

Figure~\ref{ACS} shows, as an illustration, an example of a background subtracted lightcurve observed with ACS for the solar flare of June 17, 2003. On that day GOES observed the maximum X-ray flux at 22:45~UT of a solar flare located at S08E61. The time resolution of the data shown is 1~s and the time interval extends from 30~min before the GOES peak time to 35~min after, approximately. The sudden change in the registered count rate corresponds undoubtedly to the high energy photons arrival coming from the solar flare. The prompt increase of the count rate occurs between 22:45 and 23:00~UT, in temporal correlation with the SXR flare. For energy comparison, also the background subtracted\footnote{~The background count rate for this RHESSI data has been estimated from a time interval of 5~min just before the onset of the flare.} lightcurve for three energy channels observed with RHESSI has been plotted: 25\,-\,49\,keV in red, 49\,-\,100\,keV in green and 100\,-\,250\,keV in blue. The two dashed vertical lines indicate the time interval considered to compute the counts recorded by ACS for this flare to be compared with RHESSI data. Such selection has been made in order to match the time interval considered by \inlinecite{Shih09} (see section~\ref{s4.2}). According to these authors, the two peaks occurring before the selected interval are solar emission that did not extend high enough in energy to be included in the integration time. Nevertheless, the time interval adopted for the computed observed number counts shown in Table~\ref{table1} includes these two peaks. The lightcurves of all the flares considered in this study will be available in the SEPServer database (\url{http://www.sepserver.eu}).

\section{Monte-Carlo simulations}\label{s3}

\subsection{The model}\label{s3.1}

\begin{figure}
\centering
\includegraphics[width=0.95\textwidth]{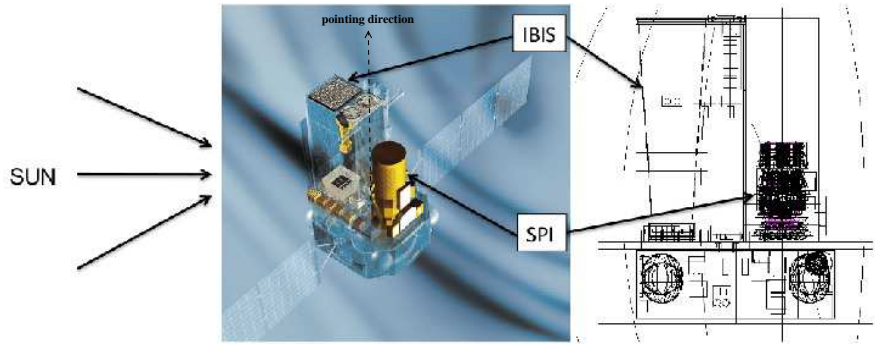}
\caption{Left: image of INTEGRAL spacecraft. SPI Ge-matrix is located inside the ACS, here represented as the yellow cylinder. Right: sketch of the INTEGRAL Mass Model and the SPI Mass Model used in the simulations. The Sun would be situated at the left side of IBIS, as indicated.}
\label{INTEGRAL}
\end{figure}

The response of ACS to incident solar high energy photons has been modelled by means of Monte-Carlo simulations, using a GEANT\footnote{~The GEANT program describes the passage of elementary particles through the matter. \url{http://wwwasd.web.cern.ch/wwwasd/geant/}}(CERN) based-code. To reproduce the instrument within the simulations, we use the spacecraft model that has previously been used for background modelling of SPI \cite{Weidenspointner03}. It contains the detailed GSFC\footnote{~Goddard Space Flight Center, NASA: \url{http://www.nasa.gov/centers/goddard/home/index.html}} SPI computer mass model \cite{Sturner03}, coupled to the INTEGRAL Mass Model (TIMM, developed at the University of Southampton, \opencite{Ferguson03}) for the other parts of the spacecraft\footnote{~An obvious error of the mass model concerning the density of the detector bench was corrected.}. Figure~\ref{INTEGRAL} shows a portrayal of INTEGRAL spacecraft (left; image from ESA, based on \opencite{Winkler03}) together with the corresponding mass model used within the simulations (right). In both cases, the Sun would be located at the left side of IBIS, in the way that the imager is always providing shadow to SPI.

The orientation of the spacecraft is determined by means of the meridional and azimuthal angles, $\theta$ and $\varphi$, respectively. The meridional angle is given with respect to the spacecraft pointing direction, with the positive values taken counter-clockwise. In this definition, a photon beam incidence completely perpendicular to the left side of IBIS (Figure~\ref{INTEGRAL}, right) would correspond to $\theta$ = 90$^{\circ}$. The azimuthal angle is defined so the Sun is considered to be located all the time around $\varphi$ =  0$^{\circ}$ (left side of IBIS).

Taking into account that solar photons can not enter to the instrument from the pointing direction of the spacecraft\footnote{~\url{http://www.isdc.unige.ch/integral/archive}}, we have extracted the orientation of INTEGRAL in space (meridional, $\theta_{\rm obs}$, and azimuthal, $\varphi_{\rm obs}$, angles) with respect to the Sun for each particular case. These computed angles, used within the Monte-Carlo code, are listed in columns [5] and [6] of Table~\ref{table1}. In accordance with the design specifications of INTEGRAL we found that the orientation of the spacecraft with respect to the Sun during the events covered the range $65^{\circ}$\,$<$\,$\theta_{\rm obs}$\,$<$\,$126^{\circ}$ in meridional angle, and $-3^{\circ}$\,$<$\,$\varphi_{\rm obs}$\,$<$\,$3^{\circ}$ in azimuth. For the October 28, 2003 solar flare, for example, we recover the angles given by \inlinecite{Gros04}, \inlinecite{Kiener06} and \inlinecite{Harris07}, $\theta_{\rm obs}$\,=\,122$^{\circ}$ and $\varphi_{\rm obs}$\,=\,0.7$^{\circ}$.

\begin{figure}[!h]
\centering
\includegraphics[width=0.65\textwidth]{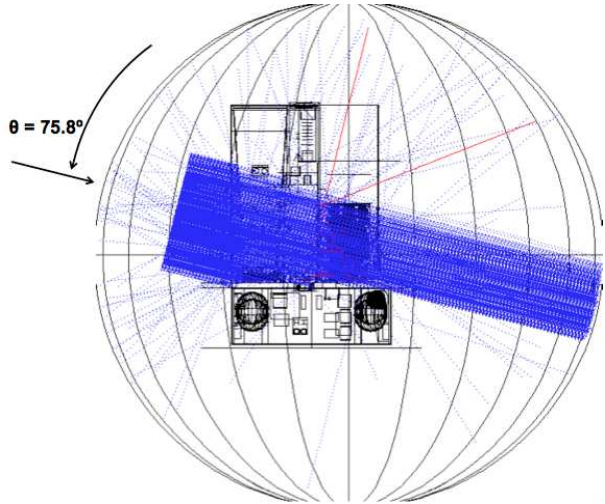}
\caption{Monte-Carlo simulation for the configuration $\theta_{\rm obs}$\,=\,75.8$^{\circ}$ and $\varphi_{\rm obs}$\,=\,-0.3$^{\circ}$ (June 17, 2003 solar flare). Dotted blue lines depict 2500 photon tracks and red dotted lines are secondary particles created by the photon interactions in the spacecraft. The arrow indicates the incoming photons direction.}
\label{simu}
\end{figure}

Figure~\ref{simu} shows as example an image of a simulation, for the configuration $\theta_{\rm obs}$\,=\,75.8$^{\circ}$ and $\varphi_{\rm obs}$\,=\,-0.3$^{\circ}$, reproducing the orientation of the spacecraft for the solar flare that occurred on June 17, 2003. The blue lines represent the track of 2500 of the simulated incoming photons, while the red lines trace the secondary particles (electrons and/or positrons) generated from interactions of the energetic photons with the spacecraft material. The arrow depicted on the left indicates the direction from which the initial photons are injected in the simulation (which would ultimately correspond to the orientation of the Sun with respect to the spacecraft for this particular case).

\subsection{Performance}\label{s3.2}

\begin{figure}[!t]
\centering
\includegraphics[width=1\textwidth]{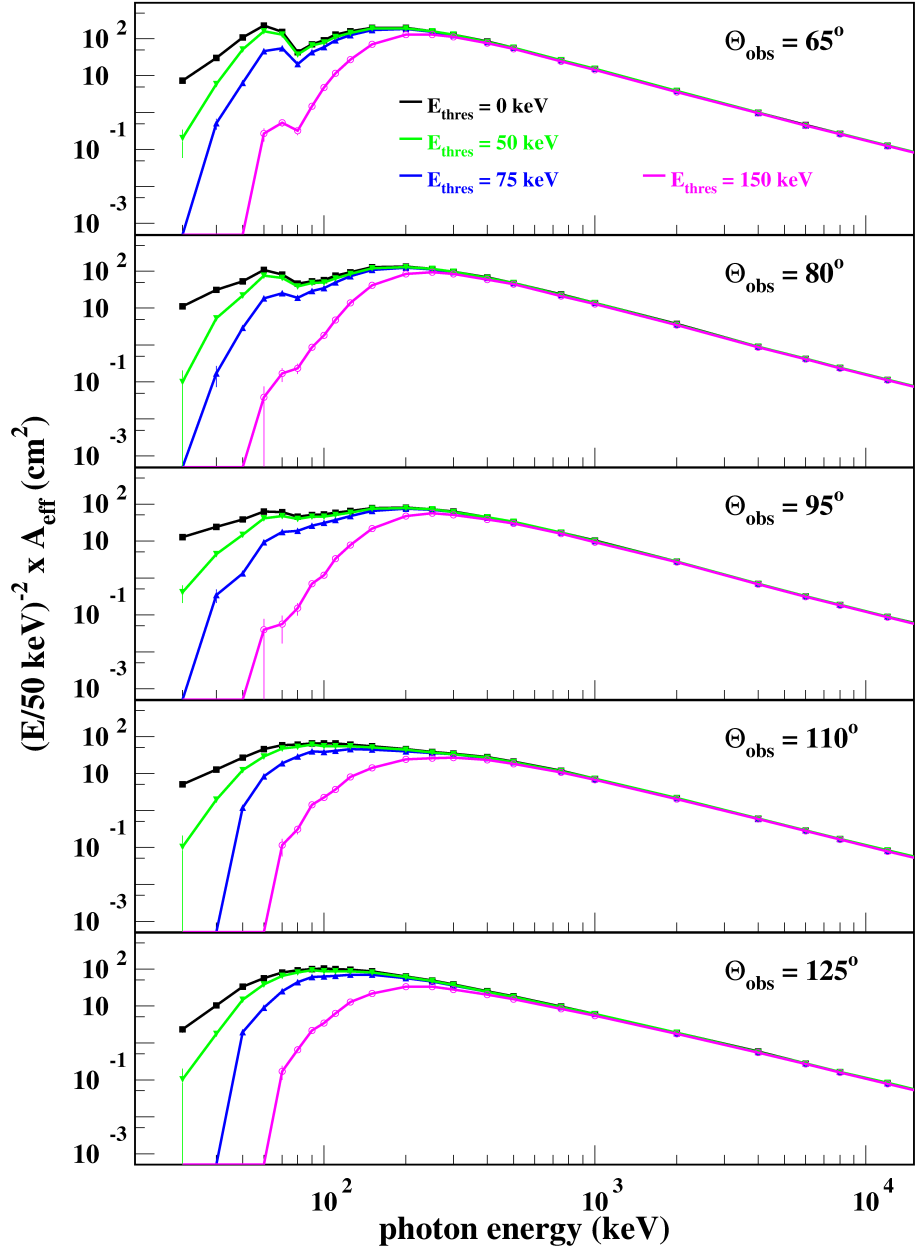}
\caption{Computed effective area as function of the energy for different orientations of the spacecraft with respect to the Sun: from top to bottom, $\theta_{\rm obs}$\,=\,65, 80, 95, 110 and 125$^{\circ}$. The electronic threshold value of the BGO detectors has been varied: in black, for a null energy threshold; green for 50~keV; blue for 75~keV and purple for 150~keV. Note that the effective area has been divided by the incoming photon energy squared, for a clearer visibility.}
\label{Aeff}
\end{figure}

To provide information about the response of the ACS/BGO detectors we have studied the dependence of the effective area ($A_{\rm eff}$) on the initial photon energy for different orientations of the spacecraft. Here, the effective area has been computed considering the initial area of the source and the efficiency of the detector. We have performed several simulations using mono-energetic beams within the range 30~keV-10~MeV. The initial beam has been considered as emitted from a disk source with a radius of $r$ = 120~cm (which accounts for a projected area of 45239~cm$^{2}$), situated at a distance $d$ = 350~cm. Such values have been chosen so the photon beam size covers completely SPI\footnote{~A larger photon beam size does not change significantly the results, the difference being smaller than 10$\%$. This error is completely negligible with respect to other uncertainties of the simulations.}. The orientation of the spacecraft has been changed, scanning the possible angle of the incoming photons. Figure~\ref{Aeff} shows, from top to bottom, the derived effective area inferred for each configuration: $\theta_{\rm obs}$\,=\,65, 80, 95, 110 and 125$^{\circ}$ (for all of them, $\varphi_{\rm obs}$\,=\,0$^{\circ}$), where $\theta_{\rm obs}$ and $\varphi_{\rm obs}$ give the orientation of the spacecraft with respect to the Sun. For any other spacecraft orientation, a direct interpolation of the values of $A_{\rm eff}$ can be done.

The electronic thresholds of the BGO detectors were individually adjusted to about 100~keV during the ground-calibration campaign of SPI with radioactive sources. A broad threshold function was estimated for the response, that results from the very modest energy resolution of the BGO detectors and the redundant cross-connection of photomultiplier tubes with pairs of crystals \cite{Attie03}. The best reproduction of ground-calibration data has been found with a mean threshold function corresponding to an energy resolution of 65~keV at 100~keV (FWHM), Kiener et al. 2013, in preparation; see also \url{http://www.sepserver.eu}). The mean energy threshold was then set, during the flight commissioning phase, to a value of 75~keV with an estimated uncertainty of $^{+75}_{-25}$~keV \cite{Kienlin03}. We have adopted these values and their uncertainties for our simulations. An energy resolution of 65\% at the threshold has been used for the threshold function. 

From Figure~\ref{Aeff} it can be clearly seen that the derived effective area largely changes depending on the orientation of the spacecraft with respect to the Sun. For $\theta_{\rm obs}$\,=\,65$^{\circ}$, $A_{\rm eff}$ may be a factor 2\,-\,3 greater than for $\theta_{\rm obs}$\,=\,125$^{\circ}$; for this latter case, the photons must cross the bottom part of the spacecraft, that results in a strong absorption of sub-MeV photons before reaching the BGO detectors. For more favorable configurations (i.e., $\theta_{\rm obs}$\,=\,65$^{\circ}$) the photons would encounter IBIS collimator in their way to SPI, which strongly absorbs photons below a few hundred keV. For further computations, here we only consider photons above 50~keV, which corresponds to a general effective threshold. The exact value of this effective energy would ultimately depend on the electronic threshold as well as on the incident angle of the photons. 

\subsection{Initial injection and event's configuration}\label{s3.3}

In order to simulate a realistic initial photon injection, series of random values for the energy $E$ following a power law distribution have been generated. Knowing that $\alpha$, the spectral index of the photon energy distribution ($f(x) = x^{-\alpha}$), may lie in the range [2;\,4] \cite{Temmer10}, we have simulated initial photon injections for three different values of $\alpha$: 2, 3 and 4. In the three cases, the initial injection was composed of 10$^{7}$ photons with energies in the range 30~keV\,$<E<$\,10~MeV. The initial source is the same disk as the one described in section~\ref{s3.2}. The orientation of the spacecraft with respect to the Sun has been set to $\theta_{\rm obs}$\,=\,75.8$^{\circ}$ and $\phi_{\rm obs}$\,=\,-0.3$^{\circ}$ (reproducing again the scenario of the June 17, 2003 solar flare). 

\begin{sidewaysfigure}
\subfloat{{\includegraphics[width=0.3\textwidth]{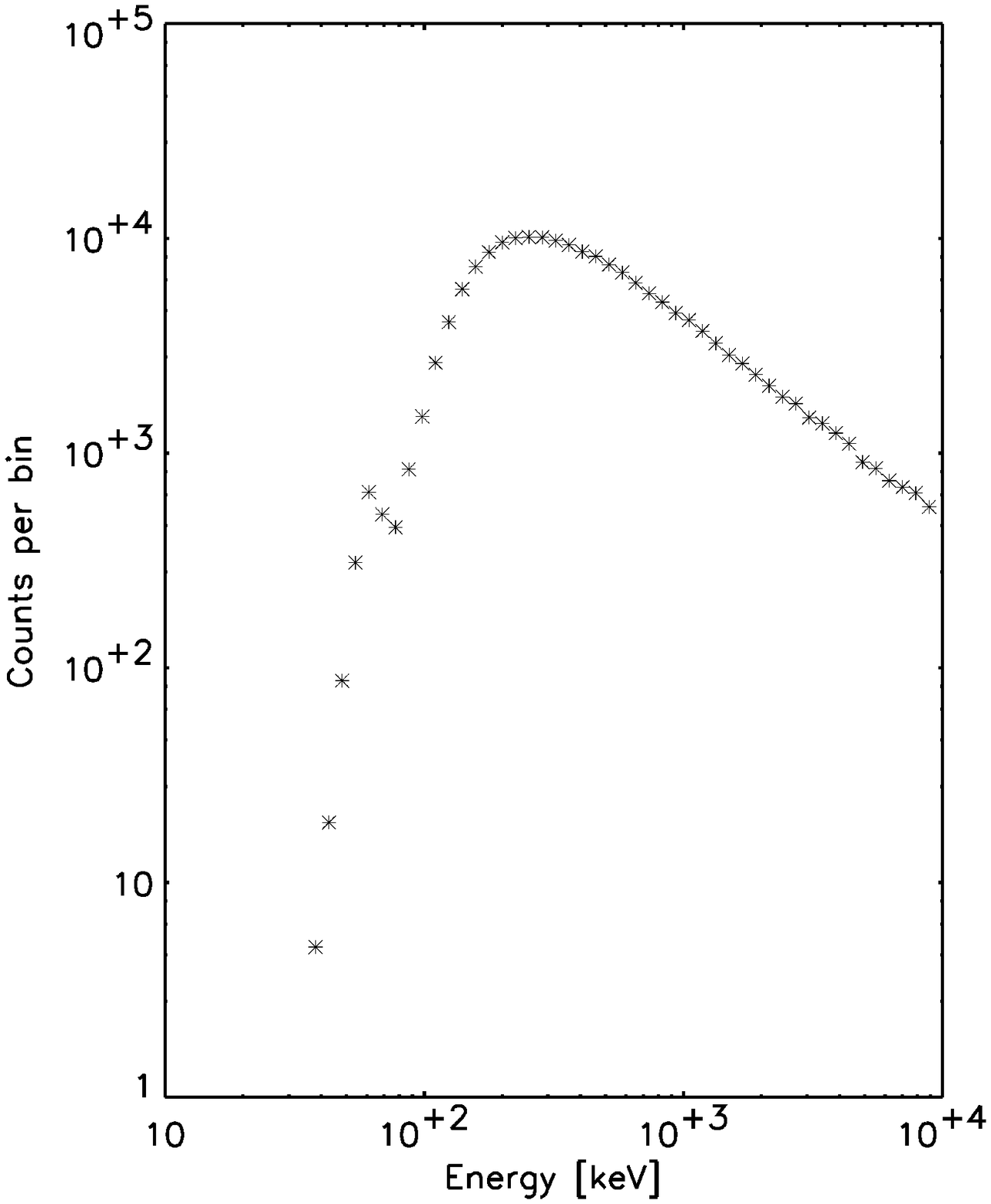}}}
\hspace{0.25cm}
\subfloat{{\includegraphics[width=0.3\textwidth]{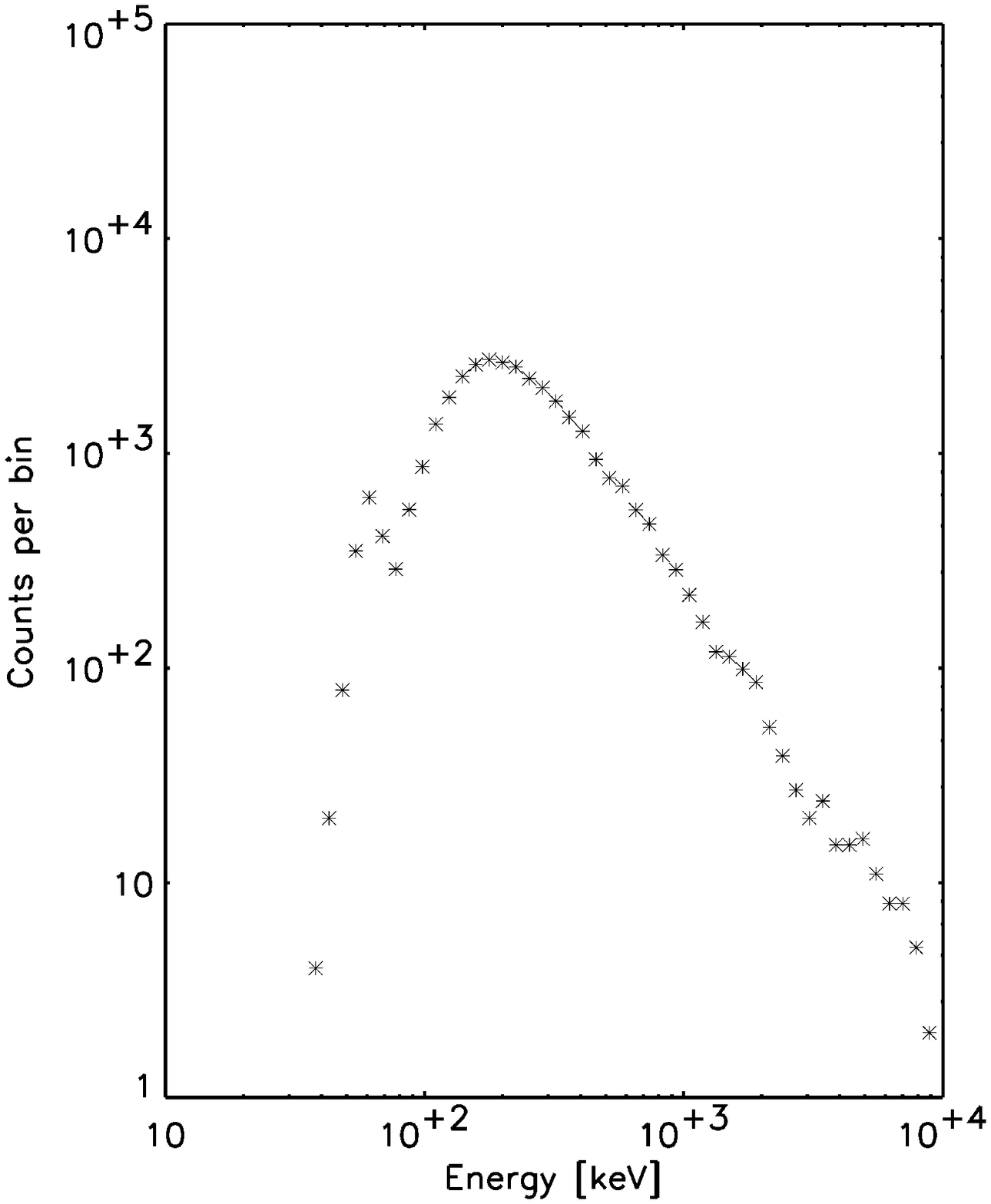}}}
\hspace{0.25cm}
\subfloat{{\includegraphics[width=0.3\textwidth]{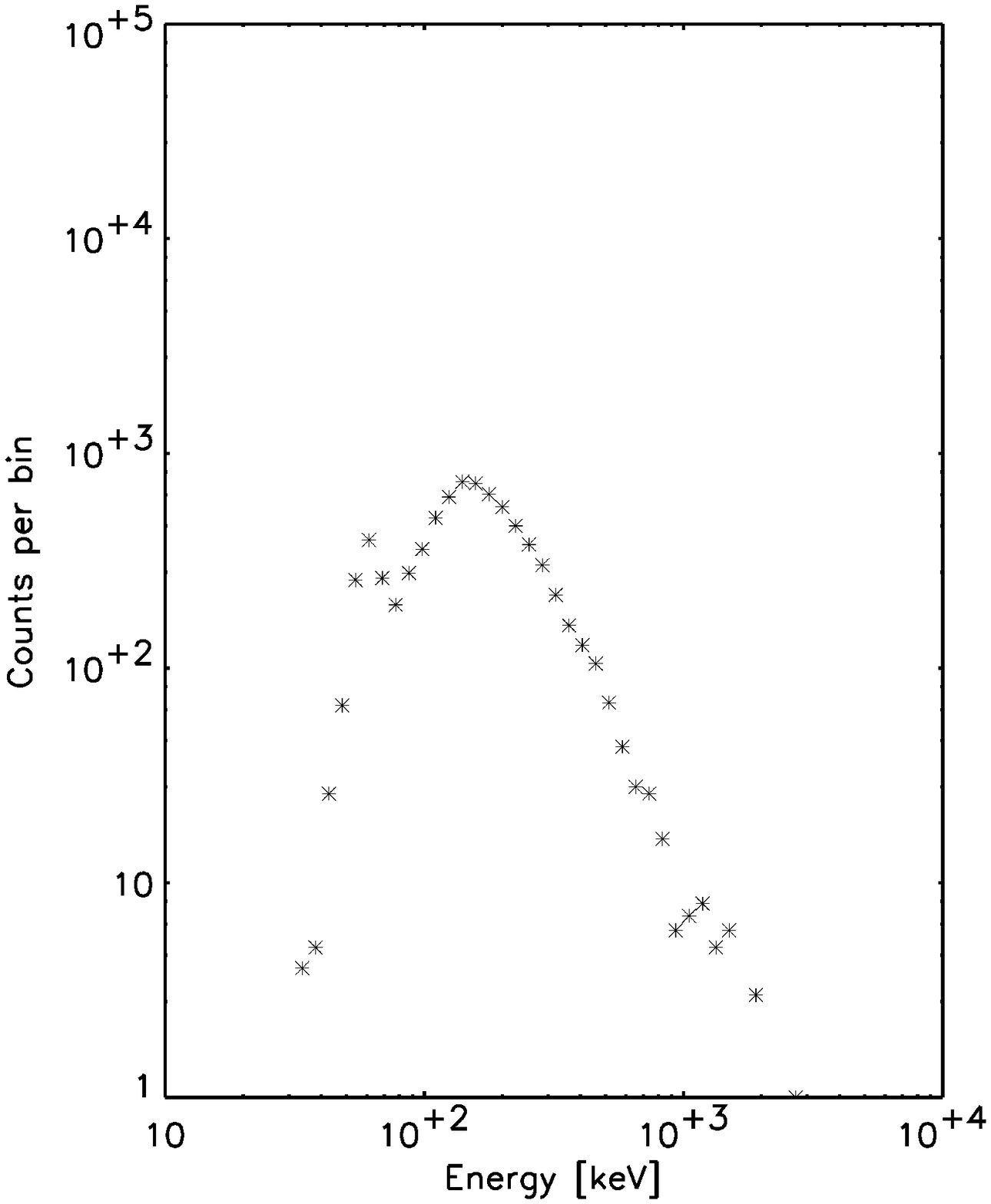}}}
\caption{From left to right, simulated response of the BGO detectors for initial power law injections with spectral index $\alpha$\,=\,2, 3 and 4, respectively, for the June 17, 2003 solar flare orientation of the spacecraft. The electronic threshold has been set to the mean value 75~keV.}
\label{injection}
\end{sidewaysfigure}

From the simulations, we can get the detector response for each case. Figure~\ref{injection} shows the results obtained for each initial injection: from left to right, for $\alpha$\,=\,2, 3 and 4, respectively. For these 3 cases, the electronic threshold of the BGO detectors has been taken fixed to the mean value 75~keV. The photons detection counting is made by means of 50 logarithmic bins. From these plots it can be concluded that for $\alpha$\,=\,2 the bulk of the hits are produced by photons with energies in the range 150\,-\,800~keV, while for  $\alpha$\,=\,3, the energy range is 100\,-\,500~keV, and for  $\alpha$\,=\,4 it decreases to 100\,-\,250~keV\footnote{~Here, the quoted energy ranges correspond to the FWHM of the count distributions.}. This result is not surprising since it is reasonable to expect more high energetic photons hitting the detector for harder initial injections. It is worthwhile to remind that ACS does not actually observe count spectra; therefore, these spectra derived from the simulations should not be misinterpreted as `predicted' count spectra.

\section{Results of the simulations}\label{s4}

\subsection{Deriving the effective area for each solar flare}\label{s4.1}

After reconstructing the spacecraft orientation with respect to the Sun for each solar flare, we have performed the response simulation of the BGO detectors for each particular case using an initial photon injection with the spectral index $\alpha$ set to 3. Then, we have computed the number of hits per bin and we have derived the corresponding effective area. Here, the effective area has been computed taking into account the initial source area and the efficiency of the detector above 50~keV; i.e., as:

\begin{equation}
A_{\rm eff} = A_{\rm ini} \frac{N_{\rm hits}}{N_{\rm ini} (> 50~\rm keV)},
\end{equation}
where $A_{\rm ini}$ is the area of the initial source (same disk as before), $N_{\rm hits}$ the number of hits recorded and $N_{\rm ini}$ ($>$ 50~keV) the number of injected photons with energies higher than 50~keV. This value just gives a general idea and we have chosen it as a realistic approximation. Nevertheless, we are acquainted that the effective energy would change from event to event, depending on the orientation of the spacecraft with respect to the Sun for each case.

\begin{figure}[!t]
\centering
\includegraphics[width=0.8\textwidth]{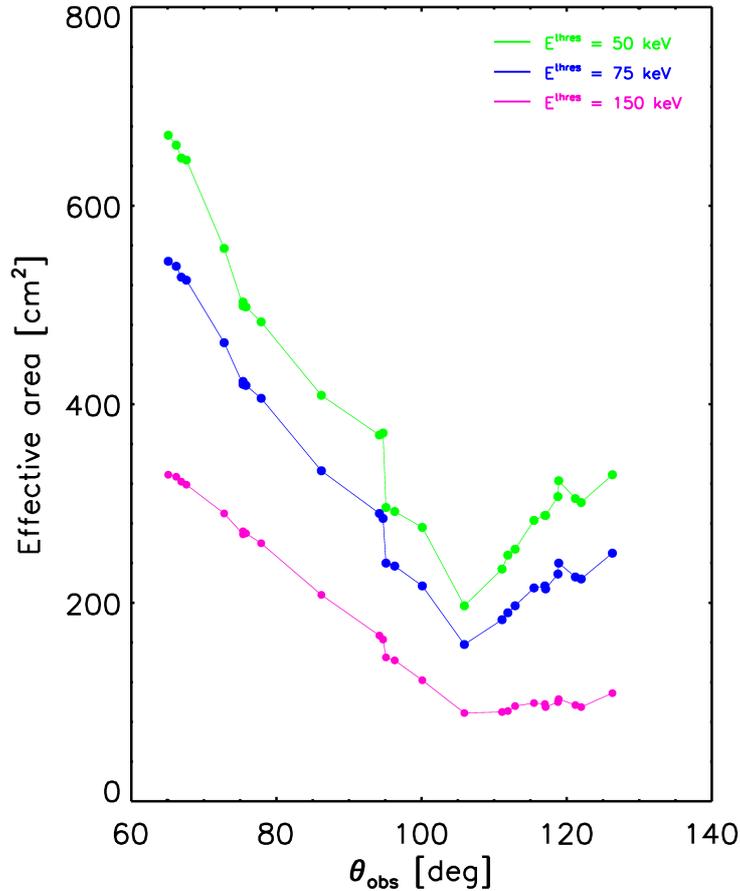}
\caption{Computed effective area as function of the incident photon angle ($\theta_{\rm obs}$) for different electronic threshold value of the BGO detectors (as labelled) and for photons with $E$ $>$ 50~keV and a initial power law with spectral index $\alpha$ = 3. The points plotted here are individual flares, including minor effects from the varying azimuthal angles ($\varphi_{\rm obs}$).}
\label{Aeff2}
\end{figure}

Column [7] of Table~\ref{table1} lists the values of the effective areas obtained when considering different electronic thresholds of the BGO detectors. The main value of the effective area given in Table~\ref{table1} corresponds to the inferred one when considering the average 91 BGO detectors threshold, 75~keV. As commented previously (section~\ref{s3.2}) such threshold value can vary from 50 up to 150~keV, which introduces the given errors in the computation of the effective area: the high area value for the lower threshold, and the low area value for the higher threshold. The variation of the effective area values with the meridional angle of incidence of the photon beam (i.e., orientation of the spacecraft with respect to the Sun), and for the different electronic thresholds of the BGO detectors is shown in Figure~\ref{Aeff2}.

From Table~\ref{table1} and Figure~\ref{Aeff2} we derive that the values of the effective area vary considerably from one flare to the other, depending basically on the orientation of the spacecraft with respect to the Sun. That comes out from the photons accessibility to the detector, according to the section of the spacecraft that the photons must cross before getting to the detector: for favorable configurations (e.g., July 6, 2006 solar flare, with $\theta_{\rm obs}$\,=\,65.1$^{\circ}$), the effective area to be seen by the spacecraft is much larger than for non favorable configurations (e.g., November 4, 2003 flare, with $\theta_{\rm obs}$\,=\,126.3$^{\circ}$). 

We can also see how the detectors' electronic threshold has a considerable influence on the effective area. The latter is much larger for the lowest electronic threshold considered (50~keV) than for the highest one (150~keV); such difference can be up to a factor of 2, because of the important contribution to the count statistics of photons with energies close to the respective electronic thresholds (see Figure~\ref{injection}).

Furthermore, to investigate how these effective areas would change with a different initial photon injection, we have performed simulations varying the spectral index, $\alpha$, as 2, 3 and 4, for three different solar flares (i.e., spacecraft orientations): June 17, 2003 ($\theta_{\rm obs}$\,=\,75.8$^{\circ}$); November 10, 2004 ($\theta_{\rm obs}$\,=\,94.7$^{\circ}$); and October 28, 2003 ($\theta_{\rm obs}$\,=\,122.0$^{\circ}$). The reason of choosing these flares is because of their different meridional angles; their azimuthal angles are similar (Table~\ref{table1}). Table~\ref{table2} presents the effective area obtained for each solar flare for the three electronic thresholds investigated, when considering different values of the initial spectral index.

\begin{table}
\caption{Comparison of the simulated effective area for three solar flares (2003 Jun 17, 2004 Nov 10 and 2005 Jan 20), considering different initial photon spectral index ($\alpha$) and different electronic threshold energies (E$_{\rm thres}$).}
\label{table2}
\begin{tabular}{ccccc}
\toprule
 & \multicolumn{3}{c}{$A_{\rm eff}$ [cm$^{2}$]} &  \multirow{2}{*}{$E_{\rm thres}$ [keV]} \\
 & $\alpha$ = 2 & $\alpha$ = 3 & $\alpha$ = 4 & \\
\hline
 \multirow{3}{*}{2003 Jun 17 ($\theta_{\rm obs}$ = 75.8$^{\circ}$)} & 149 &  498 & 236 & 50 \\
 &  1407 &  419 &  160 & 75 \\
 &  1142 &  270 &  73 & 150 \\
\cline{1-5}
 \multirow{3}{*}{2004 Nov 10 ($\theta_{\rm obs}$ = 94.7$^{\circ}$)} &  1014 &  371 &  219 & 50 \\
 & 934 &  285 &  128 & 75\\
 &  733 &  163 &  44 & 150 \\
\cline{1-5}
 \multirow{3}{*}{2003 Oct 28 ($\theta_{\rm obs}$ = 122.0$^{\circ}$)} &  691 &  301 &  193 &  50 \\
 &  615 &  224 &  115 & 75 \\
 &  429 &  95 &  29 & 150 \\
\bottomrule
\end{tabular}
\end{table}

From the results presented in Table~\ref{table2}, we see that the effective area significantly changes depending on the initial spectrum of the energetic photons. Thus, the simulated efficiency of the detector would vary for eventually different initial injections; such changes can be perfectly noticed in Figure~\ref{injection}, that illustrates the response of ACS for the case of the June 17, 2003 solar flare. As before, the effective area is largely different according to the electronic threshold considered.

Nevertheless, we can conclude that the variation of the effective area is not larger than a factor of 20\,-\,30. Then, these effective areas can be used to have a rough estimation of the solar flare fluence for photons with energies higher than 50~keV without any complementary information for the spectrum.

\subsection{Checking the model using RHESSI observations}\label{s4.2}

In order to check the performance of the model, we have derived, using RHESSI data, the number of ACS counts that we should obtain from the simulations for five solar flares (three of them coinciding with the ones described above and listed in Table~\ref{table2}). \inlinecite{Shih09} presented the RHESSI measurements of the 2.223~MeV neutron capture $\gamma$-ray line and $>$\,0.3~MeV electron bremsstrahlung continuum emissions for a set of solar flares occurred from 2002 to 2005. These emissions are indicative of $>$\,30~MeV accelerated protons and $>$\,0.3~MeV accelerated electrons, respectively. For most flares in this list, \inlinecite{Shih09} provided the integrated photon fluence for the 0.3\,-\,8.5~MeV energy range. Table~\ref{table3} presents the comparison between the observed and derived number of ACS counts, when considering different values of the electronic threshold of the BGO detectors. In our simulations, we have used the power-law indices found by the above-mentioned authors as well as the time intervals used to obtain the RHESSI fluences (A.~Shih, private communication).

\begin{table}
\caption{Comparison between the observed number of ACS counts, $N_{\rm obs}({\rm ACS})$, and the one expected from the simulations, $N_{\rm sim}({\rm ACS})$, when considering the observed RHESSI fluences, $F$($>$300~keV), for five solar flares (2003 Nov 2, 2003 Jun 17, 2004 Nov 10, 2005 Jan 17 and 2003 Oct 28, angularly ordered) and three different electronic threshold of the BGO detectors, $E_{\rm thres}$. The time intervals considered to compute both he number of ACS counts and RHESSI fluences are also given. The values of $N_{\rm sim}({\rm ACS})$ in parenthesis correspond to an enhanced contribution of nuclear $\gamma$-ray line emission (see text).}
\label{table3}
\renewcommand{\footnoterule}{}  
\begin{threeparttable}
\begin{tabular}{c|ccc}
\hline
Date                                                     & \multicolumn{3}{c}{2003 Nov 2 ($\theta_{\rm obs}$ = 75.4$^{\circ}$)} \\
\hline
time interval [hh:mm:ss]                    & \multicolumn{3}{c}{17:15:20\,-\,17:27:00}\\
$F$($>$300~keV) [ph~cm$^{2}$]  & \multicolumn{3}{c}{6500~$\pm$~15} \\
$N_{\rm obs}({\rm ACS})$                & \multicolumn{3}{c}{(5.50~$\pm$~0.1)~$\times$~10$^{7}$}\\
$E_{\rm thres}$ [keV]                        & 50     & 75     & 150\\
$N_{\rm sim}({\rm ACS})$                & 1.29 (1.36)~$\times$~10$^{8}$ &  1.07 (1.14)~$\times$~10$^{8}$ &  6.60 (7.30)~$\times$~10$^{7}$\\
\midrule

Date                                                     & \multicolumn{3}{c}{2003 Jun 17 ($\theta_{\rm obs}$ = 75.8$^{\circ}$)} \\
\hline
time interval [hh:mm:ss]                    & \multicolumn{3}{c}{22:52:40\,-\,23:04:00}\\
$F$($>$300~keV) [ph~cm$^{2}$]  & \multicolumn{3}{c}{879~$\pm$~12} \\
$N_{\rm obs}({\rm ACS})$                & \multicolumn{3}{c}{(5.45~$\pm$~0.3)~$\times$~10$^{6}$}\\
$E_{\rm thres}$ [keV]                        & 50     & 75     & 150\\
$N_{\rm sim}({\rm ACS})$                & 1.21 (1.28)~$\times$~10$^{7}$ &  1.09 (1.16)~$\times$~10$^{7}$ &  8.00 (8.06)~$\times$~10$^{6}$\\
\midrule
Date                                                     & \multicolumn{3}{c}{2004 Nov 10 ($\theta_{\rm obs}$ = 94.7$^{\circ}$)}\\
\hline
time interval [hh:mm:ss]                    & \multicolumn{3}{c}{02:08:00\,-\,02:14:00}\\
$F$($>$300~keV) [ph~cm$^{2}$]  &  \multicolumn{3}{c}{625~$\pm$~12} \\
$N_{\rm obs}({\rm ACS})$                &  \multicolumn{3}{c}{(2.67~$\pm$~0.09)~$\times$~10$^{6}$}\\
$E_{\rm thres}$ [keV]                        & 50     & 75      & 150\\
$N_{\rm sim}({\rm ACS})$                &  7.00 (7.30)~$\times$~10$^{6}$ & 5.80 (6.10)~$\times$~10$^{6}$ &  3.60 (3.90)~$\times$~10$^{6}$ \\
\midrule
Date                                                      & \multicolumn{3}{c}{2005 Jan 17 ($\theta_{\rm obs}$ = 111.9$^{\circ}$)}\\
\hline
time interval [hh:mm:ss]                    & \multicolumn{3}{c}{09:42:20\,-\,10:08:40}\\
$F$($>$300~keV) [ph~cm$^{2}$]  & \multicolumn{3}{c}{3031~$\pm$~22}\\
$N_{\rm obs}({\rm ACS})$                & \multicolumn{3}{c}{(1.0~$\pm$~0.4)~$\times$~10$^{7}$}\\
$E_{\rm thres}$ [keV]                        & 50     & 75       & 150\\
$N_{\rm sim}({\rm ACS})$                &  1.80 (1.89)~$\times$~10$^{7}$ &   1.53 (1.62)~$\times$~10$^{7}$ &  0.95 (1.04)~$\times$~10$^{7}$  \\

\midrule
Date                                                      & \multicolumn{3}{c}{2003 Oct 28 ($\theta_{\rm obs}$ = 122.0$^{\circ}$)}\\
\hline
time interval [hh:mm:ss]                    & \multicolumn{3}{c}{11:06:20\,-\,11:28:20}\\
$F$($>$300~keV) [ph~cm$^{2}$]  & \multicolumn{3}{c}{16301~$\pm$~21}\\
$N_{\rm obs}({\rm ACS})$                & \multicolumn{3}{c}{(6.0~$\pm$~0.3)~$\times$~10$^{7}$}\\
$E_{\rm thres}$ [keV]                        & 50     & 75       & 150\\
$N_{\rm sim}({\rm ACS})$                &  1.44 (1.58)~$\times$~10$^{8}$ &   1.16 (1.29)~$\times$~10$^{8}$ &  6.10 (7.40)~$\times$~10$^{7}$  \\
\hline
\end{tabular}
\end{threeparttable}
\end{table}

The number of counts recorded by ACS includes also the contribution from the nuclear $\gamma$-ray line emission. In the simulations, this contribution is estimated by taking the value of the non-corrected (of Compton-scattering by the overlying atmosphere) 2.2-MeV line fluence of \inlinecite{Shih09} and adding the 2.2-MeV line to the bremsstrahlung spectrum, which we assumed to be an unbroken power law from 50~keV to 8.5~MeV. Then, we take a template of a nuclear $\gamma$-ray line emission spectrum and we normalize it such that $F$(nuclear) = 1.1 $\times$~$F$(2.2~MeV); this value has been found for the June 12, 2010 solar flare observed by Fermi/GBM \cite{Ackermann12}. Later, we also add a 511-keV line with $F$(511~keV) = 0.5~$\times$~$F$(2.2~MeV) to the power-law spectrum \cite{Ackermann12}. The resulting spectrum is then folded with the instrument response function and yields the values for $N_{\rm sim}({\rm ACS})$ presented in Table~\ref{table3}. The values in parenthesis are for a stronger nuclear $\gamma$-ray line emission, i.e., $F$(nuclear) = 2.2~$\times$~$F$(2.2~MeV). 

From Table~\ref{table3} we can see that the number of ACS counts derived from RHESSI observations and the simulated response function of the instrument overestimates in most cases the actual number of counts observed. For the two lower thresholds the disagreement is typically a factor of 2 to 3, while the predicted counts with $E_{\rm thres} = 150$~keV approach the observed ones within 30\%. For the two flares with large meridional angles, where the solar photons cross the bottom part of the spacecraft, January 17, 2005 and October 28, 2003, the predicted values even agree within uncertainties with the observed values. The remaining difference between the latter two flares and the three flares with smaller meridional angles (where the photons cross IBIS) might be due to approximations of the mass model. However, a similar effect would probably also be observed if, for example, the electronic thresholds of the BGO detectors in bottom parts and upper parts of the shield are significantly different. Whatever the reason, we can conclude that ACS observations of solar flares agree with the expected values within 30\% for a large range of observation conditions when assuming a mean electronic threshold close to 150~keV.
 
\section{Summary and remarks}\label{s5}

We present the work on HXR/$\gamma$-ray lightcurves registered by INTEGRAL/ACS for a set of 24 solar flares of the last solar cycle that is being carried out under the FP7 SEPServer project. For all these flares, ACS recorded a substantial increase of the count rate, with a close time relation to the peak in SXR reported by some of the GOES satellites. In many cases we obtained high-quality lightcurves with 1~s resolution and, even for the weakest flares of the list with GOES SXR classes extending from M1 to X18+, all detections are mostly more than 5$\sigma$ above background, as this is one of our flare detection criteria (Section~\ref{s2}).

As high-energy solar photons come always from outside the standard observation conditions of INTEGRAL, we have performed Monte-Carlo simulations to determine the response of ACS for each flare individually. The results of the simulations clearly show an important dependence of the effective area of ACS on the angle between INTEGRAL's observation axis and the position of the Sun during the flare. For example, for a typical high-energy solar photon flux presenting a power-law spectrum in energy with index $\alpha$\,=\,3, the effective detection area varies by a factor of $\sim$~3 between the least and most favorable observation conditions.

The mean electronic threshold $E_{\rm thres}$ of the 91 BGO detectors that make up ACS has also an important effect on the effective area. After launch, $E_{\rm thres}$ was estimated to be in the 50\,-\,150~keV range which introduces an uncertainty of a factor of 2 on the actual effective area. This uncertainty could be largely reduced by comparing ACS lightcurves for flares where the photon fluxes are known from simultaneous RHESSI observations. We have shown that, for an electronic threshold value close to the upper limit of 150~keV, simulated and observed count numbers agree to within 30$\%$ for 5 different flares representing a large range of possible observation conditions. Thus, for solar HXR/$\gamma$-rays above 50~keV with power-law distribution in energy with index  $\alpha$\,=\,3, the effective area of ACS is in the range $\sim$~100\,-\,300~cm$^{2}$.

So, even if INTEGRAL was not meant to perform solar observations, ACS with its relatively large effective area may be able to record relatively faint solar flares (down to M1 class flares, perhaps lower) with a significant ($>$$5\sigma$ above background) detection, and record precise time profiles of the non-thermal emission. With the present results on the instrument response function, comparison with other satellites sensitive in a different energy range may now give an idea of the photon energy distribution. For stronger flares other instruments on board INTEGRAL, like the Ge-matrix of SPI, may also provide energy spectra. We conclude that ACS can very well be used for solar flare studies, being relevant for a better understanding of flare-accelerated particles.

\begin{acks}                                                                                                                    
This research work is supported by the European Union's Seventh Framework Programme (FP7/2007-2013) under grant agreement num. 262773 (SEPServer). The authors acknowledge the help provided from the INTEGRAL/SPI team as well as the use of the facilities of the IN2P3 computing center. We are also thankful to the SEPServer Consortium, and to A.~Shih for his kind contribution.
\end{acks}


\end{document}